\documentclass[aps,prl,groupedaddress,twocolumn]{revtex4-1}
\usepackage[dvips]{graphicx}
\usepackage{dcolumn}
\usepackage{bm}
\usepackage{amsmath,amsthm,amssymb}
\usepackage{epstopdf}

\begin{document}

\title{Spontaneous avalanche ionization of a strongly blockaded Rydberg gas}
\author{M. Robert-de-Saint-Vincent}
\author{C. S. Hofmann}
\author{H. Schempp}
\author{G. G\"unter}
\author{S. Whitlock}\email{whitlock@physi.uni-heidelberg.de}
\author{M. Weidem\"uller}\email{weidemueller@uni-heidelberg.de}
\affiliation{Physikalisches Institut, Universit\"at Heidelberg, Im Neuenheimer Feld 226, 69120 Heidelberg, Germany.}
\date{\today}

\begin{abstract}
We report the sudden and spontaneous evolution of an initially correlated gas of repulsively interacting Rydberg atoms to an ultracold plasma. Under continuous laser coupling we create a Rydberg ensemble in the strong blockade regime, which at longer times undergoes an ionization avalanche. By combining optical imaging and ion detection, we access the full information on the dynamical evolution of the system, including the rapid increase in the number of ions and a sudden depletion of the Rydberg and ground state densities. Rydberg-Rydberg interactions are observed to strongly affect the dynamics of plasma formation. Using a coupled rate-equation model to describe our data, we extract the average energy of electrons trapped in the plasma, and an effective cross-section for ionizing collisions between Rydberg atoms and atoms in low-lying states. Our results suggest that the initial correlations of the Rydberg ensemble should persist through the avalanche. This would provide the means to overcome disorder-induced-heating, and offer a route to enter new strongly-coupled regimes. 
\end{abstract}

\maketitle

Ultracold plasmas (UCPs) formed by photo-ionizing ultracold neutral atomic or molecular gases~\cite{killian1999,killian2007} offer an ideal laboratory setting to better understand exotic phases of matter such as dense astrophysical plasmas~\cite{horn1991} and laser induced plasmas \cite{Remington1999}. 
Experimental and theoretical progress is driven by the potential to reach the so-called strongly-coupled regime~\cite{ichimaru1982,killian2007b}, in which the Coulomb interaction energy dominates over the kinetic energy of the ions, giving rise to collective effects and strong spatial correlations between particles. It is quantified by the coupling parameter $\Gamma = q^2/4 \pi \epsilon_0 a k_B T \gg 1$, where $a$ is the Wigner-Seitz radius, $q$ the electron charge, and $T$ the ion temperature.
In laser-cooled gases, $\Gamma\approx 0.1 - 2$ is readily achieved, which has allowed the observation and driving  of collective mechanical modes\,\cite{Fletcher2006, *twedt2012, *castro2010}. 
Reaching deep into the strongly-coupled regime, however, has remained out of reach partly due to disorder-induced-heating (DIH), where the Coulomb interaction energy due to the initially random distribution of the atoms is converted into kinetic energy of the ions~\cite{murillo2001, pohl2004, *bergeson2011, *simien2004, *cummings2005}.

Gases of atoms in high-lying (Rydberg) states offer an alternative approach to study UCPs, as they can be easily ionized, and the strong Rydberg-Rydberg interactions lead to dramatic new effects. 
The spontaneous evolution of an attractively interacting Rydberg gas into an UCP has been observed in several experiments\,\cite{robinson2000,*walz-flannigan2004, *morrison2008,*heidemann2007, li2004}, which initiated in-depth studies on the ionization mechanisms~\cite{Beterov2007, Amthor2007, tanner2008} and on the electron-ion recombination dynamics towards Rydberg states\,\cite{killian2001, *robicheaux2002, *fletcher2007, *saquet2011}. 
Recently\,\cite{weber2012}, the long-term formation of an ionic cloud from attractive and repulsive Rydberg states has been observed, and its back-action onto Rydberg laser-excitation rates has been characterized.
At the high densities achievable using optical or magnetic traps, the Rydberg blockade effect~\cite{Comparat2010} gives rise to strong correlations in the initial gas~\cite{Robicheaux2005}. Since these correlations resemble those in strongly coupled plasmas~\cite{gericke2003}, Rydberg blockade should help mitigate DIH and provide a path to new strongly coupled regimes\,\cite{murillo2001, kuzmin2002}. However,  mechanical collapse driven by attractive forces\,\cite{Amthor2007, Pillet2009} 
and anisotropic interactions could destroy the correlations.

 \begin{figure}
  \begin{center}
      \includegraphics[width=0.9\columnwidth]{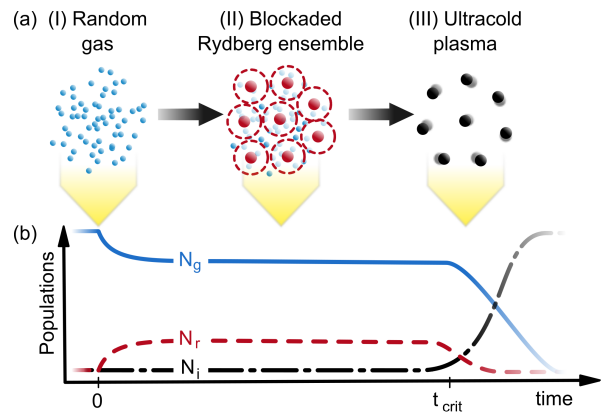}
      \caption{(Color online) \textbf{Generation of an ultracold plasma from a Rydberg-blockaded gas.} (a) An initially uncorrelated gas of ultracold atoms is prepared in an optical dipole trap (I). By coherent laser coupling we excite some atoms to a Rydberg state. The strong Rydberg-Rydberg interactions prevent the excitation of close-by pairs, leading to spatial correlations (II). After a time $t_{\rm{crit}}$ the Rydberg gas is observed to spontaneously ionize into an ultracold plasma (III). (b) Qualitative dynamics of the involved populations: ground state atoms $N_g$ (solid line), Rydberg atoms $N_r$ (dashed line), and ions $N_i$ (dash-dotted line). 
 }
    \label{fig:phyics_idea}
  \end{center}
\end{figure}
In this letter, we report on the formation dynamics of an ultracold plasma from an initially spatially-correlated gas of \emph{isotropic} and \emph{repulsively} interacting Rydberg atoms in the blockade regime. Despite the relative stability of repulsively interacting Rydberg ensembles, we observe, after a well defined but controllable time, a spontaneous avalanche ionization, as evidenced by a sudden increase in the number of detected ions, accompanied by a depletion of the ground and Rydberg state populations. We identify the relevant processes leading to plasma formation, and using a simple coupled rate-equation model we quantify the relevant formation rates. We observe that repulsive interactions delay the onset of avalanche ionization, and lead to a density dependence which differs significantly from previous observations~\cite{vitrant1982, li2004}. We estimate that the typical time-scale for plasma formation is short compared to the motional dynamics of the ions, suggesting that the initial correlations should be preserved in the plasma phase.

The basic principle behind our observations can be understood in three main stages (Fig.~\ref{fig:phyics_idea}). (I) An initially randomly distributed gas of ultracold neutral atoms is excited to Rydberg states via a continuous two-photon laser coupling. (II) Due to the Rydberg blockade, each Rydberg-excited atom blocks further excitations within a radius $R_c$ leading to density-density correlations which resemble those of a gas of hard-spheres. 
After a short time, the Rydberg density reaches steady state, however over time Rydberg atoms start to decay by a combination of blackbody photoionization~\cite{Beterov2007} and ionizing collisions with atoms in the ground and intermediate states \cite{barbier1987, *kumar1999}, which leads to a gradual increase in the number of charged particles in the system. (III) Once a critical number of ions $N_{\rm{crit}}$ accumulates the resulting space charge can trap subsequently produced electrons.  At this critical time $t_{\rm{crit}}$, rapid electron-Rydberg collisions trigger an ionization avalanche, leading to the formation of an UCP. Since the avalanche is triggered by the fast moving electrons the plasma forms rapidly compared to the motion of the ions. Therefore the original spatially ordered structure of the Rydberg ensemble should be preserved. This is predicted to reduce the effects of DIH by several orders of magnitude~\cite{murillo2001}. Finally, after this sudden ionization of most of the Rydberg ensemble, the continuous laser excitation feeds the avalanche until the ground state population is fully depleted.

In our experiments we measure simultaneously the atom number in the ground or intermediate state $N_g$, the Rydberg atom number $N_r$, and the ion number $N_i$ in a small excitation volume through a combination of absorption imaging and ionization detection. Our experimental setup is as follows: dense atomic samples of up to $1.5 \times 10^5$  $^{87}$Rb atoms in the $|5S_{1/2}, F=2\rangle$ state are loaded into a single beam optical dipole trap. We hold the atoms in the trap for a short time to equilibrate at a temperature of $T = 230 \pm 30\,\rm{\mu K}$. We then release the atoms, and let the cloud expand during a typical time-of-flight of $200\,\rm{\mu s}$ to a radially symmetric Gaussian distribution with $e^{-1/2}$-radii of $\sigma_{\rm{radial}} = 28 \pm 1 \rm{\mu m}$ and $\sigma_{\rm{long}} = 310 \pm 10 \rm{\mu m}$. The peak atomic density is varied between $n_0 = (1 - 10)\times 10^{10}\,\rm{cm}^{-3}$ while maintaining an approximately constant cloud shape by reducing the overall atom number during trap loading. 

Rydberg atoms in the $|55S_{1/2}\rangle$ state are excited using a doubly resonant narrow-band two-photon excitation via the intermediate $|5P_{3/2}\rangle$ state. Both lasers are aligned perpendicular to the symmetry axis of the cloud. The first excitation step is realized with a nearly uniform beam at $780\,\rm{nm}$ wavelength with intensity $I_p = 2.9 \,\rm{mW/cm^2}$. The second laser, at 480~nm, is aligned counter propagating to the first and is focused onto the center of the cloud with a Gaussian waist $w=40\,\rm{\mu m}$ and a peak coupling strength $\Omega/2\pi = 5.8\,\rm{MHz}$. The excitation volume $V$, defined by the intersection of the cloud and coupling beam, is approximated by a cylinder along the coupling beam, of length $L = \sqrt{2 \pi} \sigma_{\rm{radial}} \approx 70 \,\mu \rm{m}$ and radius $R = w/\sqrt{2} \approx 28\, \mu \rm{m}$.
To perform the experiment we simultaneously pulse on both lasers for times $t$ ranging from 0 to 38$\,\rm{\mu s}$. During this time, $V$ expands by less than $\sim 20\,\%$. To stop the excitation, we switch on an electric field which field ionizes the Rydberg atoms and subsequently guides all ions to a micro-channel plate detector (MCP). 

\begin{figure}
  \begin{center}
      \includegraphics[width=0.85\columnwidth]{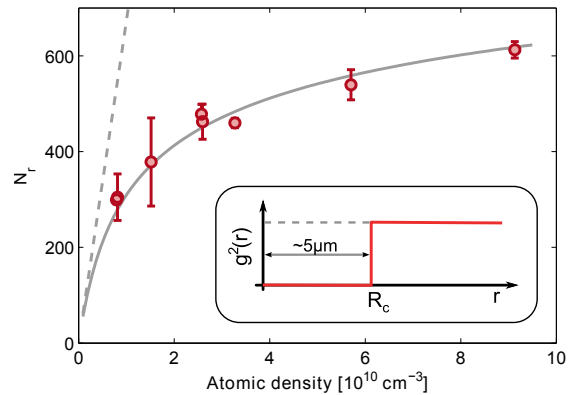}
      \caption{(Color online) \textbf{Density dependence of the Rydberg number for short times.}
The atoms in $|55S\rangle$ are field-ionized after $3.7 \mu s$ of continuous laser coupling. Strong interactions lead to a Rydberg excitation suppression for increasing ground state densities, in contrast to the interaction--free case (dashed line). The data is in good agreement with a hard sphere model (solid line) with a sphere diameter corresponding to the blockade radius $R_c \approx 5\,\mu m$. Inset shows the spatial correlation function typical for a Rydberg blockaded gas\,\cite{Robicheaux2005}. }
    \label{fig:blockade}
  \end{center}
\end{figure}

We first investigate the dependence of the Rydberg population on the density $n_0$ for a short excitation time $t = 3.7\,\rm{\mu s}$ (Fig.\,\ref{fig:blockade}), at which we can rule out the influence of spontaneously formed ions. We observe the characteristic nonlinear dependence and saturation of $N_r$ which is a consequence of Rydberg blockade~\cite{Singer2004,*Tong2004}. The observed scaling is well described using a classical hard sphere model in the steady-state (solid line in Fig.\,\ref{fig:blockade})\cite{Ates2011note}. In this model, each Rydberg-excited atom is assumed to produce an exclusion sphere of radius $R_{\rm{c}}\approx\sqrt[6]{(2 C_6 \Gamma)/\Omega^2}$ which reduces the fraction of ground state atoms available for subsequent excitation. From the strength of the van-der-Waals interactions, $C_6/2\pi \approx 50~\rm{GHz~\mu m^6}$ for the $|55S_{1/2}\rangle$ state, and from the intermediate-state decay rate, $\Gamma/2\pi \approx 6.1~\rm{MHz}$, we get $R_{\rm{c}} \approx 5\,\rm{\mu m}$. We assume independent excitation $A\times\phi$ and de-excitation rates $B$. Interactions are accounted for through the fraction of available ground-state atoms $\phi\leq 1$, which decreases for increasing Rydberg density and is estimated using the Carnahan-Starling expression for hard spheres (see Supplemental Material and\,\cite{talbot1991}). The ratio $A/(A+B)= 0.38$ is constrained by the steady state solution to the (single-atom) three-level optical Bloch equations for our experimental parameters. Results of the analytic model were verified using Monte-Carlo simulations of the excitation process. 

From this model we derive the steady state $N_r$ as a function of $n_0$ assuming homogeneous density and coupling strength in $V$. As seen in Fig.~\ref{fig:blockade} the hard-sphere model is in excellent agreement with the data, with the overall detection efficiency $\sim 0.4$ as only adjustable parameter. From  comparison to the data we find that interactions suppress $N_r$ by up to a factor of 8. At the highest densities the fraction of blockaded atoms $1-\phi \approx 0.9$, and the packing fraction $\eta=\pi N_{\rm{r}} R_{\rm{c}}^3/6V \approx 0.2$, which indicates the presence of strong correlations in the system. 
 
 \begin{figure}
  \begin{center}
      \includegraphics[width=0.9\columnwidth]{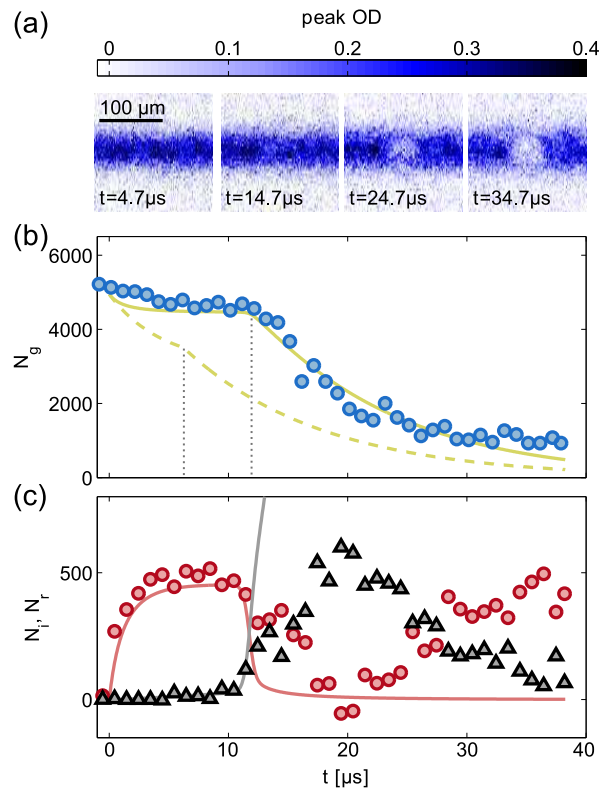}
      \caption{(Color online) \textbf{Measurement of the complete avalanche ionization dynamics for $N_g$, $N_r$ and $N_i$.}  (a) Absorption images taken after different durations of continuous laser coupling on a cloud of initial density $n_0 = 2.8 \times 10^{10}\rm{cm}^{-3}$. (b) Number of ground-state atoms $N_g$ in the excitation volume. The data are compared to a rate equation model (see text), accounting for (solid line) or neglecting (dashed line) the Rydberg blockade mechanism. Vertical lines highlight the avalanche onset time. (c) Simultaneously recorded Rydberg atom population $N_r$ (circles) and ion population $N_i$ (triangles). }
    \label{fig:exp_finding}
  \end{center}
\end{figure}

To study the evolution of the system we probe the ground state atomic distribution using resonant absorption imaging. This is performed using a second light pulse on the lower transition, applied $1\,\mu s$ after the end of the Rydberg excitation. Figure\,\ref{fig:exp_finding}a shows a sequence of absorption images of the center of the cloud, recorded for different excitation times, at a density $n_0 = 2.8 \times 10^{10}\rm{cm}^{-3}$. For times longer than $\approx 10~\rm{\mu s}$ we observe the appearance of a hole coinciding with the position of the focused 480\,nm excitation laser. Over $10-20~\mu$s the hole grows until the cloud becomes locally transparent. We estimate the remaining number of atoms $N_g$ in $V$ (shown in Fig.\,\ref{fig:exp_finding}b) by fitting to the images the difference of two peak functions. 
The time-dependence shows initially a slow decrease of $N_g$, until a critical time $t_{\rm{crit}}\approx 12~\rm{\mu s}$ at which point starts a rapid depletion. 

This behaviour is a signature of the ionization avalanche of the Rydberg population. To independently measure $N_r$ and $N_i$, we perform two experiments: in the first, we use an electric field above the Rydberg ionization threshold, and we detect both Rydberg atoms and ions using the MCP. In a second experiment, we switch a field below the ionization threshold, which has a negligible effect on the Rydberg atoms but is sufficient to guide the ions to the detector. After relative calibration of the detection efficiencies for both experiments, we substract the two signals to deduce $N_r$ and $N_i$ individually (Fig.~\ref{fig:exp_finding}\,c). $N_r$ reaches a quasi-steady-state value $\approx 500$ after $\approx 4\rm{\mu s}$. Around $t_{\rm{crit}}\approx 12\,\mu$s, the number of ions reaches $N_i=120$ and starts to increase rapidly. This number is already sufficient to trap electrons, thus leading to the formation of an UCP\,\cite{killian2007b}, with lifetime $\approx 9~\rm{\mu s}$. Applying a small electric field during excitation suppresses the avalanche ionization, thereby confirming that the plasma formation is related to an accumulation of charges.

The observed dynamics can be described by using a simple coupled-rate equation model which incorporates the Rydberg blockade effect, seed ionization processes, and an avalanche term driven by ionizing electron-Rydberg collisions (see Supplemental Material). We neglect the role of Rydberg-ion interactions, assuming the net charge imbalance is relatively small and the ions are effectively screened by the mobile electrons. This assumption is consistent with our experimental observation that the Rydberg resonance does not significantly shift after plasma formation, but only becomes broadened\,\cite{weber2012}. On the observed timescale, the repulsive $|55S_{1/2}\rangle$ atoms are not expected to undergo ionizing Rydberg-Rydberg collisions at a sufficient rate to seed the avalanche \cite{amthor2007b}. Instead, we attribute the seed processes to a combination of blackbody photoionization~\cite{Beterov2007} and ionizing collisions either with ground-state atoms or with the relatively large population in the intermediate state $|5P_{3/2}\rangle$~\cite{barbier1987, *kumar1999}. Once $N_i$ exceeds $N_{\rm{crit}}$, we consider $N_e = N_i - N_{\rm{crit}}$ electrons to be trapped in a plasma, which subsequently interact with the remaining Rydberg population.

We fit the model results simultaneously to a set of ground state depletion curves, with initial atomic densities ranging from $(0.8-9)\times 10^{10}\rm{cm}^{-3}$. The black-body photoionization rate is taken from ref.~\cite{Beterov2007}. This leaves three free parameters that we extract from the model: the overall Rydberg excitation rate $A$ from the ground and intermediate states, the effective seed collisional ionization cross section $\sigma_{col}$ \footnote{This cross-section encapsulates ionizing collisions from atoms in the ground or first excited state regardless of their state, thus containing contributions from different ionizing mechanisms~\cite{barbier1987, kumar1999}.}, and the average energy of trapped electrons $E_e$, which constrains both $N_{\rm{crit}}$ and the non-linear rate for electron-Rydberg collisions $\gamma_{av}$. 

Typical predictions of the model are shown in Fig.~\ref{fig:exp_finding}b-c. The sharp evolution of $N_g$ is well reproduced, although the model does not completely reproduce $N_r$ and $N_i$ after the avalanche. 
We attribute this to three-body recombination towards Rydberg states counted in the detection process, and to plasma expansion, both not included in the model. The best fit parameters and uncertainties are: $A = (8.7 \pm 0.3) \times 10^{-2} \, \rm{MHz}$;  $\sigma_{\rm{col}} = 0.73 \pm 0.21\,\mu \rm{m}^2$, a factor of 8-15 larger than the geometrical cross-section of the Rydberg atoms ($\sigma_{geo} \approx \pi n^{\star 4} a_0^2 \approx 0.064 \,\mu \rm{m}^2$, where $a_0$ is the Bohr radius and $n^{\star}$ is the effective principal quantum number); and $E_e = (3.0\, \pm 0.9)\times 10^{-3}\,$eV, i.e. approximately 60$\%$ of the electron binding energy in $|55S_{1/2}\rangle$, similarly to\,\cite{li2004}. Given our geometry, $E_e$ relates to $N_{\rm{crit}} = 26 \pm 7$ (typically 5$\,\%$ of the initial Rydberg number), and to $\gamma_{\rm{av}} \approx \sigma_{geo} \sqrt{E_e /m_e}\approx (1500 \pm 200)\,\rm{MHz}\cdot\mu m^{3}$~\cite{vitrant1982}.

\begin{figure}
  \begin{center}
      \includegraphics[width=0.9\columnwidth]{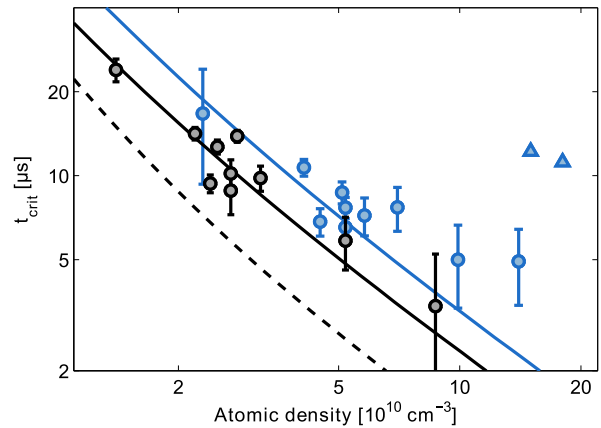}
      \caption{(Color online) \textbf{Density dependence of the ionization avalanche onset time
	       $t_{\rm{crit}}$} (log-log scale). 
Dark and light symbols correspond to experiments with excitation volumes $V$ and $V_{\rm{reduced}} \approx V/2$, respectively.  Circles and triangles correspond to the $|55S_{1/2}\rangle$ and $|40S_{1/2}\rangle$ states, respectively. 
Continuous lines are the model predictions for $|55S_{1/2}\rangle$ for both volumes, using the same parameters fitted on experiments in $V$. The dashed line shows the model outcome in $V$ when neglecting the blockade effect.}
    \label{fig:density_dependence}
  \end{center}
\end{figure}

To investigate the regimes in which a plasma develops, we extracted $t_{\rm{crit}}$ from the data as a function of density (Fig.~\ref{fig:density_dependence}). 
We observe a critical dependence, following an approximate power law of $t_{\rm{crit}} \propto n_0^{-\alpha}$ with $\alpha=1.0\pm0.1$ on the considered density range. 
We have also studied the role of the excitation volume by performing experiments for a shorter time-of-flight of $100~\mu$s, which leads to a reduced volume $V_{\rm{reduced}} \approx V/2$. We observe that the onset times are delayed, since $N_{\rm{crit}}$ decreases slower with $V$ than the total atom number\,\cite{killian2007b}. This effect is well reproduced by the model, when accounting for the new geometry with conserved rates and cross sections. 

From the model we conclude that the Rydberg blockade has a significant effect on the formation of the UCP, delaying $t_{\rm{crit}}$ by typically 80\,$\%$ for the data shown in Fig.~\ref{fig:density_dependence}. In such a time, Rydberg interactions are suspected to develop strong spatial correlations~\cite{Robicheaux2005}. 
Our model predicts a sudden avalanche ionization of the complete Rydberg ensemble within $\approx 2 \, \mu$s. Within this time, the ions only move by at most 600\,nm, which is much less than their average spacing ($\gtrsim R_c$): thus, the correlations from the Rydberg sample are most-likely preserved in the avalanche.
The observed packing fraction $\eta \approx 0.2$ is similar to the highest considered in \cite{murillo2001}: provided the laser couplings are turned off immediately after $t_{\rm{crit}}$, we expect that the importance of DIH as compared to final ion interactions should be strongly reduced. 

To further study the effect of Rydberg interactions on the avalanche dynamics, we compared the dynamics from different Rydberg states, while keeping all other experimental parameters (in particular $\Omega$) approximately constant. In comparison to $|55S_{1/2}\rangle$, the $|40S_{1/2}\rangle$ state appears more stable, exhibiting plasma formation in $V_{\rm{reduced}}$ only above densities of $n_0\approx 1\times 10^{11}$ (triangles in Fig.~\ref{fig:density_dependence}). 
Rescaling the parameters $\sigma_{col}, \gamma_{av} \propto n^{\star 4}$; $R_c \propto n^{\star 11/6}$; and $E_e \propto 1/ n^{\star 2}$ \cite{li2004}, the model accounts for approximately half of the relative shift in $t_{\rm{crit}}$. This discrepancy could be due to a different scaling of the ionization cross-section $\sigma_{col}$. For the $|55D_{3/2}\rangle$ state we observe an immediate production of ions ($t_{\rm{crit}}\rightarrow 0$), presumably due to the anisotropy of D-state interactions which breaks the blockade effect and leads to an increased rate of ionizing Rydberg--Rydberg collisions \cite{Amthor2007}.  

In conclusion, we have experimentally investigated the spontaneous evolution from a strongly blockaded and spatially correlated Rydberg gas to an UCP. Simultaneous absorption imaging and ion detection allow us to observe the complete dynamics of this process. In particular we identified a regime in which the ionization avalanche is faster than the motional timescale of the produced ions, suggesting a new route to the production of strongly coupled UCPs with large ionic coupling parameters $\Gamma$. In addition, Rydberg states allow for a large tuning of the correlation length given by $R_c \propto n^{\star 11/6}$. Future experiments could directly resolve the spatial structure of the resultant plasmas, for example by using atomic species with more than one valence electron~\cite{McQuillen2012}, or by using Rydberg states as a 
sensitive probe~\cite{gunter2012,*olmos2011}. This would open promising new avenues to experimentally investigate novel effects in UCPs, such as liquid-vapor phase coexistences, critical points~\cite{shukla2011} and strongly coupled liquid phases.  


\acknowledgements{This work is supported in part by the Heidelberg Center for Quantum Dynamics and
the Deutsche Forschungsgemeinschaft under WE2661/10.2; M.R.D.S.V. (grant number FP7-PEOPLE-2011-IEF-300870) and  S.W. (grant number PERG08-GA-2010-277017) acknowledge support from the EU Marie-Curie program.}


\section{Supplemental material}
\subsection{Coupled rate-equation model of the avalanche dynamics}
We describe the avalanche ionization by a coupled rate-equation model between the number of atoms in
the ground or intermediate state, both included in $N_g$, the number of Rydberg atoms $N_r$, the
number of ions $N_i$, and the number of electrons $N_e$ trapped in the plasma. Only two coupled
differential equations describe the evolution of the system:
\begin{eqnarray}
\frac{dN_g}{dt} &=& -A \phi N_g + B N_r \\
\frac{dN_r}{dt} &=& A \phi N_g - B N_r - \gamma_{bbi} N_r  \\
             &~& - \gamma_{col} N_r N_g / V   - \gamma_{av} N_r N_e /V\,. \nonumber
\end{eqnarray}
Here, $V$ is the excitation volume and  $\gamma_{bbi}$ $\approx
150 /s $ is the black-body photo-ionization rate~\cite{Beterov2007}.

We deduce $N_i$ and $N_e$ from the conservation of the total particle number $N_{tot}$, 
neglecting expansion of the plasma. Considering the electron trapping
criterion~\cite{killian2007b}, we get:
\begin{eqnarray}
N_i &=& N_{tot} - N_g - N_r \\
N_e &=& N_i - N_{\rm{crit}}  ~~ \rm{,if}>0\,.
\end{eqnarray}
The fraction $\phi$ of ground state atoms that, despite the blockade mechanism, is available for
Rydberg excitation, is evaluated following the Carnahan-Starling approximation for
hard-spheres~\cite{talbot1991}:
\begin{equation}
\phi = \exp{ \left(\frac{- 8 \eta + 9 \eta^2 - 3 \eta^3}{(1-\eta)^3}\right)}\,,
\end{equation}
where $\eta = \pi N_r R_c^3 / 6V$ is the packing fraction of hard-core spheres with radius $R_c/2$.
We relate the non-linear seed ionization rate $\gamma_{col}$ to an effective collision cross section $\sigma_{col}$ between Rydberg atoms and atoms in either the ground or the intermediate state:
\begin{equation}
\gamma_{col} = \sigma_{col}  \sqrt{\frac{16 k_B T}{\pi m_{Rb}}}\,,
\end{equation}
where $T$ is the temperature of the atomic cloud, $k_B$ is the Boltzmann constant, and $m_{Rb}$ is
the mass of $^{87}$Rb.

The non-linear avalanche rate $\gamma_{av}$ is related to the electron energy $E_e$ by
\begin{equation}
\gamma_{av} = \sigma_{geo} \sqrt{E_e / m_e}\,.
\end{equation}
Here $\sigma_{geo} = \pi n^{\star 4} a_0^2$ is the geometric cross-section of Rydberg atoms
\cite{vitrant1982}, with $n^{\star}$ the effective quantum number and $a_0$ the Bohr radius. The
electron mass is given by $m_e$ and we assume that half of $E_e$ is of kinetic form.

Finally, assuming $V$ to be a homogeneous cylinder of length $L$ and radius $R$, we evaluate the
critical ion number for electrostatic trapping of electrons with total energy $E_e$:
\begin{equation}
N_{\rm{crit}} = \frac{8 E_e L \pi R^2 \epsilon_0}{q^2 \left[L (-L + \sqrt{L^2 + 4 R^2}) +
   4 R^2 \rm{csch}^{-1}(2 R/L)\right]}\,,
\end{equation}
where $\epsilon_0$ is the vacuum dielectric permittivity, $q$ is the electron charge, and
$\rm{csch}^{-1}$ is the inverse hyperbolic cosecant function.



\end{document}